\documentstyle[12pt]{article}
\def\lsim{\mathrel{\rlap{\raise 2.5pt \hbox{$<$}}\lower 2.5pt}}
\def\gsim{\mathrel{\rlap{\raise 2.5pt \hbox{$>$}}\lower 2.5pt}}
\begin{document}
\thispagestyle{empty}
\begin{flushright}
UNIL-TP-2/95, hep-ph/9503490\\
\end{flushright}

\begin{center}
{\bf{\Large Testing the $\Delta S=\Delta Q$ Rule
with Exclusive Semi-Leptonic Kaon Decays$^*$}}
\vskip 0.5cm
B. Ananthanarayan\\
Institut de physique th\'eorique, Universit\'e de Lausanne,\\ [-2mm]
CH 1015, Lausanne, Switzerland.\\
\vskip 1.cm
\end{center}
\begin{abstract}
We consider the possibility of violations
of the selection rule $\Delta S=\Delta Q$
at an appreciable level in {\it exclusive}
semi-leptonic decays of Kaons.  At $\Phi$-Factories, intense
Kaon beams will be available and will probe among others,
the semi-leptonic decays $K_{l4}$ and $K_{l3\gamma}$ in addition to $K_{l3}$
and could provide novel testing grounds for the $\Delta S=\Delta Q$ rule.
In particular, the branching ratio of $K_{l3\gamma}$ is non-negligible
and could be used to probe new phenomena associated with the
violation of this selection rule.  Furthermore, we  modify
certain di-lepton event rate ratios and asymmetries and
time asymmetries that have
been constructed by Dass and Sarma for di-lepton events from
Beon decays to test the $\Delta B=\Delta Q$ at the $\Upsilon (4S)$,
to the Kaon system at the $\phi(1020)$.
We find that the large width of the $K_S$ relative to that of
$K_L$ plays an important role in enhancing some of the
time asymmetries.
\end{abstract}
\noindent{\underline{\hspace{11.6cm}}}\\
\begin{small}
$^{*}$Work Supported by the Swiss National Science Foundation
\end{small}
\newpage

Intense Kaon beams will soon be available with the commissioning of
$\Phi$-Factories including DA$\Phi$NE which is expected to
begin running as early as 1996\cite{Maiani1}.  Indeed, at
DA$\Phi$NE many high precision experiements will attempt to
test many well-accepted principles, including the validity of
the CPT theorem\cite{Maiani2}.  A violation of this would
require a drastic reappraisal of what is considered well-understood.

A less dramatic but important discovery would be violation of
the  $\Delta S=\Delta Q$ rule in Kaon decays
at a level significantly larger
than predicted by the standard model.
In the standard model, processes that violate the $\Delta S=\Delta Q$
rule arise only at the two-loop level\cite{Dib} in the decay of Kaons.
This rule also attains significance in that all tagged Kaon experiments
will identify a lepton in the final state\cite{Cline}.
Furthermore
important timing information will also be available on the
Kaon decays which will render the measurements of time asymmetries
feasible for this system\cite{Cline,Hayakawa}.
Studies have been
performed for the violation
of the rule in the $K_{l3}$\cite{Hayakawa,Dambrosio}
decay which has a  substantially
large branching ratio.

 At DA$\Phi$NE, however, important
information will be gathered on the semi-leptonic decays
$K_{l4}$ and $K_{l3\gamma}$ as well, since these provide
important tests to
the predictions of and information on parameters of the chiral expansion
at one-loop\cite{Bijnens}.
Despite their relatively low branching ratios, sufficiently
large number of events will be seen which
affords the opportunity of considering $exclusive$ semi-leptonic
decays of the Kaons in some detail.
Here we propose that violations of
the  $\Delta S=\Delta Q$ rule be searched for in such decays
of the Kaons in addition to the $K_{l3}$ decays.
This possibility has not been considered in any detail
to the best of our knowledge.  It may therefore be fair
to say that the possibility of the
violation of the $\Delta S=\Delta Q$ rule in these
additional channels is essentially unconstrained by data.
Finally, the availability
of more than one channel offers the advantage that we can
construct asymmetries that are odd under the exchange of
the exclusive final states that are labeled by $i$ and $j$
since the Kaons are produced in a $P$-wave at $\phi(1020)$.

Such an approach has been taken in case of the Beon system
at the $\Upsilon(4S)$ by Dass and Sarma\cite{Dass}.
In the case of Beons, at
the outset it is necessary to consider exclusive
semi-leptonic decays since
they have comparable branching ratios.  Specific asymmetries
were constructed from like and unlike sign di-lepton event rates
and time asymmetries were constructed
 in order to exploit this information.  In
Ref.\cite{Dass}, CPT invariance has been assumed throughout and
the violation of the $\Delta B=\Delta Q$ rule has been considered
in some generality where the parameter $r_i$ that provides a measure of
this is assumed to be complex and independent for each of the
exclusive channels $i$.  Here we show that these asymmetries can
be modified for DA$\Phi$NE in a very straightforward way
and provide an ideal set of quantities that are essentially
new as regards the Kaon system.
We compute some of them numerically for the exclusive channels
of interest to us.  The specific characteristics of each of the
Kaon and Beon systems often interchanges the role of some of
the asymmetries in terms of their capacity to probe the violations of
the $\Delta F=\Delta Q$ rule where the flavor $F$ stands for $S$ in
the Kaon and $B$ in the Beon system.
While some of the results of the present investigations
are encouraging, we remark here
that this approach is probably only one of several that are available.
Furthermore,
the considerations of say Ref.\cite{Hayakawa,Dambrosio}
may be applied to the $K_{l4}$ and $K_{l3\gamma}$ decays as well.

We begin by considering the Kaon system in the standard manner
by writing down the $K_S$ and $K_L$ in terms of the flavor eigenstates
$K_0$ and $\overline{K}_0$ as:
\begin{eqnarray}
|K_S\rangle = p |K_0\rangle + q |\overline{K}_0\rangle \nonumber \\
|K_L\rangle = p |K_0\rangle - q |\overline{K}_0\rangle
\end{eqnarray}
with $(|p|^2+|q|^2)=1$.

In the presence of CP violation, we have $g(\equiv q/p)\neq 1$
and is defined in terms of $\epsilon$ (
neglecting $\epsilon'$) as $g=(1-\epsilon)/(1+\epsilon)$.
We also have the definitions $x=(m_2-m_1)/\Gamma$, $y=(\Gamma_2-
\Gamma_1)/2\Gamma$, $\Gamma=(\Gamma_1+\Gamma_2)/2$ and
$a=(1-y^2)/(1+x^2)$, where $m_1,\ m_2, \ \Gamma_1$ and $\Gamma_2$
are the masses and widths of the two physical eigenstates
$K_S$ and $K_L$ respectively.

These parameters are accurately determined for the Kaon system
(in contrast to that of the Beon system) and we have for the
central values of these parameters from \cite{Montanet}:
$m_2-m_1=3.51\cdot 10^{-12}\ MeV$, $\Gamma_1=7.37\cdot 10^{-12}\
MeV$, $\Gamma_2=1.27\cdot 10^{-14}\ MeV$, $\Gamma=3.69\cdot 10^{-12}\
MeV$, $x=0.95$, $y=-0.997$, $a=3.6\cdot 10^{-3}$,
$|\epsilon|=2.3\cdot 10^{-3}$, $\phi(\epsilon)=44^0$
and $|g|=0.997$.  [It is important
to note that in the Beon system, although the parameters are not
very well determined, due to the large phase space available for
the decay of the Beon, $y$ is expected to be small and correspondingly
$a$ is expected to be relatively large; this is clearly interchanged
in the present system.]

Following the notation of Ref.\cite{Dass} with the obvious
replacement of the flavor $B$ everywhere by the flavor $S$,
we write down the $\Delta S=\Delta Q$ conserving amplitudes:
\begin{eqnarray}
A_i=\langle i l^+ |T|K_0\rangle, \ \overline{A}_i=\langle \tilde{i} l^-
|T|\overline{K}_0\rangle
\end{eqnarray}
(the tilde denotes the CPT conjugate state)
and those violating the rule as:
\begin{eqnarray}
\rho_iA_i=\langle i l^+ |T|\overline{K}_0\rangle, \
\overline{\rho}_i\overline{A}_i=\langle \tilde{i} l^-
|T|K_0\rangle
\end{eqnarray}
Furthermore, since we assume the validity of CPT everywhere, we
have the relations:
\begin{eqnarray}
\overline{A}_i=A_i^*,\ \overline{\rho}_i=\rho_i^*
\end{eqnarray}
where the $\rho_i$ and $A_i$ are in general complex.
We also have the definition $r_i\equiv g\rho_i$ and
$\overline{r_i}=(\overline{\rho_i}/g)$.

The label $i$ denotes any of the exclusive semi-leptonic
decay channels.  For definiteness we consider only the three
decay channels $K_{l3}$, $K_{l4}$ and $K_{l3\gamma}$ and in
all the considerations that follow we have the following
correspondences between the label $i$ and the state:

(1) $i=1$ standing for the state $|\pi^+ \nu_l\rangle$
and describes the $K_{l3}$ decay,

(2) $i=2$ standing for the state $|\pi^+ \pi^0
\nu_l\rangle$ and describes the $K_{l4}$ decay,

\noindent and

(3) $i=3$ standing for the state $|\pi^+ \nu_l\gamma\rangle$
and describes the $K_{l3\gamma}$ decay.

The branching ratios for these processes are predicted in the chiral
expansion at one-loop and are expected to be confronted at DA$\Phi$NE.
For the purposes of the discussion that follows we will consider only
the central values for them and a serious consideration of the errors
will have to be taken up at a later state when the analysis of the
data is undertaken.  The three numbers of interest may
be obtained from the detailed considerations of chiral
perturbation theory\cite{Bijnens,Bijnens2} for the
rare decays and from the Particle Data Group
for the $l3$ decays \cite{Montanet} which lists the following
branching ratios :
\begin{eqnarray}
BR(K_{\mu 3}^0)=2.7\cdot 10^{-1},\ BR(K_{e 3}^0)=3.9\cdot 10^{-1} \nonumber \\
BR(K_{\mu 4}^0)=4.5\cdot 10^{-6}, \ BR(K^{0}_{e4})=4.7\cdot 10^{-5} \nonumber\\
BR(K_{\mu 3 \gamma}^0)
=5.6\cdot 10^{-4}, \ BR(K^{0}_{e3\gamma})=3.8\cdot 10^{-3}
\end{eqnarray}

We will define the allied combinations $h_{ij}$ and $g_{ij}$
as the ratios:
\begin{eqnarray}
h_{ij}={{|A_i|^2}\over{|A_i|^2+|A_j|^2}} \nonumber\\
g_{ij}={{|A_iA_j|}\over{|A_i|^2+|A_j|^2}}
\end{eqnarray}
for $i,j=1,2,3$ ($h_{ii}=g_{ii}=1/2$).

As in Ref.\cite{Dass} in much of the following discussion we will
parametrize the effects of the non-standard transitions in terms of
the parameters $r_i\equiv g \rho_i$ and $\overline{r}_i=\overline{\rho}_i/g$.
Furthermore, all the asymmetries and quantities we discuss here
are identical with those of \cite{Dass}.  We give
only those definitions we consider necessary to make the
presentation simpler and the reader is
often refered
to Ref.\cite{Dass} for detailed discussion in order to avoid repetition.

We begin by first computing the exclusive di-lepton ratio
$\chi_{ij}$\cite{Dass} defined as the relative number of like-sign
dilepton events all of which originate in $i$ and $j$ or
$\tilde{i}$ and $\tilde{j}$:
\begin{eqnarray}
& \displaystyle
\chi_{ij}={{N_{ij}^{++}+N_{ij}^{--}}\over{N_{ij}^{++}+N_{ij}^{--}+
N_{ij}^{-+}+N_{ij}^{+-}}} &\nonumber \\
& \displaystyle
=\{ 1+4a({\rm Im}\langle r\rangle _{ij})^2+{{4a}\over{(1-a)}}
\left(g_{ij}|r_i-r_j|\right)^2 \}\chi &
\end{eqnarray}
where
\begin{equation}
\chi={{(1-a)(1+|g|^4)}\over{(1-a)(1+|g|^4)+2(1+a)|g|^2)}}=0.4982
\end{equation}
and
$r_{ij}$ is the weighted average given by:
\begin{equation}
r_{ij}=h_{ij}r_i+h_{ji}r_j.
\end{equation}

The final expression for $\chi_{ij}$ therefore is:
\begin{equation}
\chi_{ij}=\left\{1+0.0144\left(({\rm Im}\langle r\rangle _{ij})^2+
\left(g_{ij}|r_i-r_j|\right)^2\right)\right\}(0.4982)
\end{equation}
The asymmetry for a single channel $\chi_{ii}$ is given by:
\begin{equation}
\chi_{ii}=(1+0.0144\ ({\rm Im}\ r_i)^2)(0.4982)
\end{equation}

We have also used here and in what follows the approximation:\\
$|1-r_i\overline{r_j}|^2\simeq (1-2{\rm Re} r_i \overline{r_j}^*)$.
Here we add that the factor $4a(\simeq
{{4a}\over{1-a}})=0.0144$ and is relatively small.

The same-sign di-lepton asymmetry $\alpha_{ij}$,
constructed from events that emerge from the two channels
$i$ and $j$ is computed next:
\begin{eqnarray}
& \displaystyle
\alpha_{ij}={{N_{ij}^{++}-N_{ij}^{--}}\over{N_{ij}^{++}+N_{ij}^{--}}}= &
\nonumber \\
& \displaystyle{1-|g|^4\over 1+|g|^4}\left\{1+{{4}\over{1+|g|^4}}
\left[{\rm Re}(\langle r \rangle_{ij})^2-{{1+a}\over{1-a}}(g_{ij}
|r_i- r_j|^2)\right]\right\} & \nonumber \\
& \displaystyle
=0.0066\left\{1+2.013\left[{\rm Re}(\langle r \rangle_{ij})^2-0.99\
(g_{ij}|r_i- r_j|^2)\right]\right\} &
\end{eqnarray}

Both the observables $\chi_{ij}$ and $\alpha_{ij}$ suffer from the
disadvantage of being quadaratic in the $r_i$ and any deviations from
the standard model predictions are likely to be small for these,
indeed as anticipated in Ref.\cite{Dass}.

The time-asymmetries constructed out of like-sign dilepton
events of \cite{Dass} may be considered next for the
case at hand.  These are obtained by straightforward
replacements of the quantities that specify the Kaon system
and the final forms are given below:
\begin{eqnarray}
{\cal A}_{l^{+(-)} l^{+(-)}}(ij)=\mp 0.068\ {\rm Im}(r_j-r_j)+
1.986\ {\rm Re}(r_i-r_j)
\end{eqnarray}
where ${\cal A}_{l^+ l^+}(ij)$ is defined as the asymmetry:
\begin{equation}
{{[\nu(ij)-\nu(ji)]}\over{[\nu(ij)+\nu(ji)]}}
\end{equation}
and $\nu(ij)$ stands for for the number of dilepton events in which
the $l^+$ associated with the channel $i$ occurs $earlier$ than
the one associated with $j$.  The associated symbol for $l^-$,
 ${\cal A}_{l^- l^-}(ij)$
is obtained by replacing $(i,j)$ by $(\tilde{i},\tilde {j})$.

We note here that the asymmetries above merit serious consideration
since they are linear in the differences of the parameters $r_i$
and terms in these expressions are not met with necessarily small
coefficients.  In particular, it may be possible that at
DA$\Phi$NE the $K_{l3\gamma}$ decay can be used in conjunction with
$K_{l3}$ decay, given the formers non-negligible branching ratio.

We then have the final expression for the unlike-sign dilepton events
the definition of which is easily obtained from eq. (19)
and eq. (20) of Ref.\cite{Dass}
\begin{equation}
{\cal A}_{l^+ l^-}(ij)=0.0136\ {\rm Im}\langle r \rangle_{ij}.
\end{equation}

Finally we present the channel asymmetry given in eq. (21)
of Ref.\cite{Dass} that is given by:
\begin{equation}
{\cal A}_{l^+ l^-}(i\tilde{j}+\tilde{i} j)=
{{-2y}\over{(1+a)}} {\rm Re} (r_i-r_j)= 1.986 \ {\rm Re} (r_i-r_j)
\end{equation}
is relatively interesting for the case at hand since $|y|$ is
a quantity that is almost 1.  This is one example of a quantity
that is perhaps less interesting in the  Beon system but significant
in the Kaon system.

In conclusion, we note that the availability of intense Kaon beams
opens prospects of probing violations of the selection rule
$\Delta S=\Delta Q$ in the semi-leptonic of Kaons not only to
higher precision but in novel ways.  This includes looking in the
exclusive Kaon decay channels provided by the $K_{l4}$ and
(more so) in the $K_{l3\gamma}$ decays due to their non-negligible
branching ratios and by constructing asymmetries out of di-lepton
event samples whose exclusive final states are known.  An example
is provided by adapting the asymmetries of Dass and Sarma constructed
for the Beon system at the $\Upsilon(4S)$ to the present system,
{\it viz.} the Kaon system at the $\phi(1020)$.  Asymmetries that
are linear and odd under the exchange of final states are particularly
useful since the Kaons are produced here in a $P$-wave.  Some of
them are shown to be sensitive to the selection rule violation due
to the fundamental role played by the large width of the $K_S$
relative to that of the $K_L$.

\bigskip

\noindent{\bf Acknowledgements:}  It is a pleasure to thank
K. V. L. Sarma for encouraging discussions and a critical reading of the
manuscript.  We thank G. V. Dass for useful remarks,
L. M. Sehgal for a clarifying correspondence and
J. Gasser and K. Sridhar for conversations.

\end{document}